\documentclass[fleqn,12pt,twoside]{article}
\usepackage{espcrc1}
\usepackage{graphicx}
\usepackage[figuresright]{rotating}

\newcommand{\AmS}{{\protect\the\textfont2
  A\kern-.1667em\lower.5ex\hbox{M}\kern-.125emS}}

\title{Beam--Charge Asymmetry associated with DVCS at HERMES}

\author{F. Ellinghaus\address[MCSD]{DESY Zeuthen, Platanenallee 6,
         15738 Zeuthen, Germany}%
        \thanks{E-mail: Frank.Ellinghaus@desy.de}
        \\ (On behalf of the HERMES Collaboration)
        }
       
\begin{document}

\maketitle

\begin{abstract}
We report the first observation of an azimuthal asymmetry 
in the hard electroproduction of real photons with 
respect to the charge of the incoming lepton beam.
The asymmetry is attributed to the interference between the 
Bethe--Heitler process and the deeply--virtual Compton scattering  
process, 
which gives access to the latter at the amplitude level.
This process appears 
to be the theoretically cleanest way
to access generalized parton distributions.
The data have been accumulated by the HERMES experiment at DESY,
scattering the HERA 27.6 GeV electron/positron beam
off an unpolarized hydrogen gas target. 
\end{abstract}

\section{Introduction}
Inclusive and semi--inclusive deep--inelastic scattering (DIS) 
is and has been extensively used to study the internal structure
of the nucleon.
Recent theoretical progress 
is mostly related to
exclusive reactions and their description in terms of generalized
parton distributions (GPDs), also referred to as off--forward
or skewed parton distributions \cite{Mue94,Rad97,Ji97}.
This theoretical framework 
takes into account the dynamical correlations between partons
of different momenta in the nucleon.
The well--known parton distribution functions and form factors 
turn out to be the limiting cases and moments of GPDs.
Of particular interest is the second moment of two 
unpolarized quark GPDs, which for the first time offers
a possibility to determine the total angular momentum carried
by the quark in the nucleon \cite{Ji97}.
Very recent theoretical ideas indicate that GDPs might
be able to 
describe correlations between the longitudinal
and transverse structure of the nucleon \cite{Bur,Die02}.

\section{Deeply--Virtual Compton Scattering}
The theoretically cleanest way to access GPDs appears to be the
deeply--virtual Compton scattering (DVCS) process, i.e. the   
hard exclusive leptoproduction of real photons with the target
nucleon remaining intact. Since the Bethe--Heitler (BH) process 
has an identical final state and therefore the amplitudes of both processes
add coherently, 
the interference between them can be used to access the DVCS amplitudes.
The leading--order and leading--twist interference term \cite{Die97}
\begin{equation} \label {Diehl}
I = \pm \frac {4 \, \sqrt 2 \, m \, e^6}{ t \,  Q \, x_{Bj}} 
\frac{1}{\sqrt {1-x_{Bj}}} \times 
[\cos \phi \frac{1}{\sqrt {\epsilon (\epsilon - 1)}}
\mathrm{Re} \tilde M^{1,1} 
- P_l \sin \phi \sqrt { \frac {1 + \epsilon}{\epsilon}} 
\mathrm{Im} \tilde M^{1,1}]
\end{equation}
depends on the charge and the helicity of the incident lepton, 
where +(-) denotes a negatively (positively) charged lepton
with polarization $P_l$. 
Note that terms suppressed by $O(1/Q)$ and involving higher
$\phi$--moments, were omitted in the above equation.
Here m represents the proton mass, t the square of the four--momentum 
transfer to the target, $-Q^2$ the virtual--photon four--momentum squared
and $\epsilon$ is the 
polarization parameter of the virtual photon.
Measuring the dependence of certain cross section asymmetries on the 
azimuthal angle $\phi$, defined as the angle between the lepton 
scattering plane and the photon production plane, 
provides information about the 
DVCS amplitude combination $\tilde M^{1,1}$
which can be expressed in terms of GPDs.
The extraction of an asymmetry with respect to the beam spin, accessing the imaginary part 
of $\tilde M^{1,1}$, has already been carried out by 
the HERMES experiment \cite {HER_DVCS}.
The extraction of a beam--charge asymmetry, accessing the real part
of $\tilde M^{1,1}$, is described in the following.

\section{The Beam--Charge Asymmetry}
The data have been accumulated with the HERMES spectrometer
\cite {HER_Spec} at DESY during the 1998 (2000) 
running periods. 
The HERA 27.6 GeV electron (positron) beam was scattered off 
an unpolarized hydrogen gas target.
Events were selected if they contained exactly one photon and
one charged track, identified as the scattered electron (positron) 
and fulfilling the kinematical requirements $Q^2 > 1$~GeV$^2$, 
$W^2 > 4$~GeV$^2$ and $\nu < 23$~GeV. Here $W$ denotes the 
photon--nucleon invariant mass and $\nu$ is the virtual--photon energy.
The angle between the real and the virtual photon was required to be 
within 15 and 70~mrad.

\begin{figure} [h!]  
\begin{center}
\includegraphics[width=0.6\textwidth, height=7.6cm]{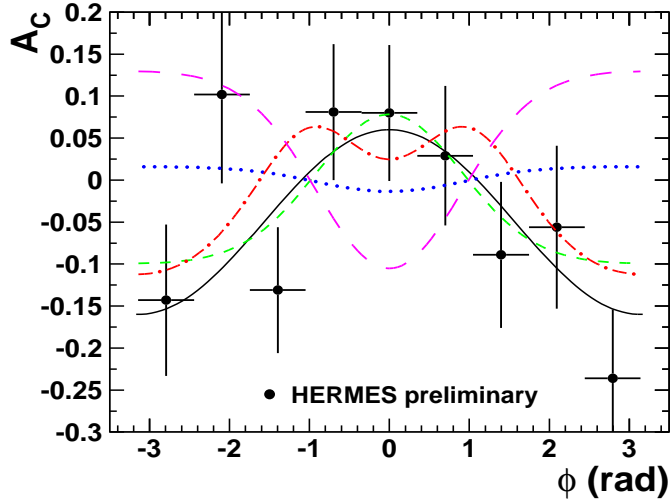}
\caption{Beam--charge asymmetry for the hard electroproduction
of photons as a function of the azimuthal angle $\phi$.
The data correspond to the missing mass region 
between $-1.5$~GeV and $+1.7$~GeV.
The solid line represents a $P_1 + P_2 \cos (\phi)$ fit to the data.
The other curves are model calculations \cite{Van02} 
described in the text.}
\label{toll}
\end{center}
\end{figure}
In figure \ref{toll} the azimuthal dependence of the extracted 
beam--charge asymmetry (BCA) 
\begin{equation} \label {bca}
A_C (\phi) = \frac {N^+ (\phi) - N^- (\phi)}{N^+ (\phi) + N^- (\phi)}
\end{equation} 
is shown, where $N^+$ ($N^-$) represent the luminosity--normalized yields
for the corresponding beam charges.
Using only events with a missing mass $M_x$
between -1.5~GeV and +1.7~GeV, the data show the
expected $\cos \phi$--behavior as can be seen by the solid line
which represents a $P_1 + P_2 \cos (\phi)$ fit.
The missing mass is defined as 
$M_x = \sqrt {(q + P_p - k)^2}$
with $q$, $P_p$ and $k$ being the four--momenta of the 
virtual photon, the target nucleon and the real photon,
respectively. Negative values of the missing mass are due
to the finite momentum resolution of the spectrometer,
in which case $M_x = - \sqrt {-M_x^2}$ was defined.
The beam used at HERMES is polarized; to minimize its effects, 
positron (electron) data sets were selected to have 
opposite polarizations $P_l$ ($-P_l$)
to eliminate the $\sin \phi$--dependence
in the numerator of equation (\ref {bca}).
This dependence still exists in the denominator but should be negligible
compared to the size of the BH amplitudes.
The fit reveals a positive $\cos \phi$ amplitude with a value of 
$0.11 \pm 0.04$ (stat) and an offset of $P_1 = -0.05 \pm 0.03$ (stat).
The latter value could possibly be related to a constant
term in the interference, omitted in equation (\ref {Diehl}) since it appears
at twist--3 \cite {bel_muel}.
However, the sign and the size of the amplitude might already
contain some valuable information as can be seen by a comparison 
to model calculations \cite {Van02} which were carried out in a kinematic
regime ($<Q^2>$ = 2.5~GeV$^2$, $<x_{Bj}>$ = 0.11, $<t>$ = -0.25~GeV$^2$) 
close to the HERMES kinematics
($<Q^2>$ = 2.8~GeV$^2$, $<x_{Bj}>$ = 0.12, $<t>$ = -0.27~GeV$^2$).
The calculations are the same as described in reference \cite {Van00}.
Certain GPDs are parameterized assuming that the t--dependent part factorizes,
while the t--independent part is modelled by a two--component form,
using the so--called double--distribution formalism \cite{rady} completed 
by the D--term \cite {poly_weis}. 
The D--term is related to the previously unstudied ``stress tensor'' of
hadronic matter.
It contributes to the real part of the 
DVCS amplitude only and therefore can be investigated with the BCA for the first time. 
Calculating the D--term contribution in the chiral quark soliton model results in
the dashed (twist--2) and the dashed--dotted line (twist--3) in figure \ref{toll}. 
The twist--2 result without the additional D--term is 
represented by the dotted line. 
The long--spaced dashed line is 
the twist--2 DVCS charge asymmetry when reversing the sign of the D--term.
Thus the preliminary HERMES data already favor the existence of the D--term
whereby, if it exists, the data clearly determine its sign.
However, very recent calculations including twist--3 effects, but using a different 
GPD model without D--term contribution describe the HERMES data as well \cite {Mue02}.

\begin{figure} [h!]
\begin{center}
\includegraphics[width=0.55\textwidth, height=7.7cm]{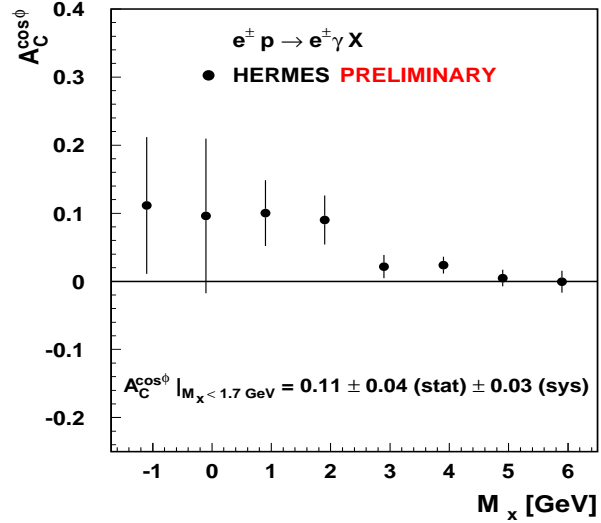}
\caption{Difference between the positron and the electron $\cos \phi$--weighted 
moments versus the missing mass $M_x$.}
\label{toller}
\end{center}
\end{figure}
To be able to compare the result in the exclusive region with that in the non--exclusive
region, the difference between the $\cos \phi$--weighted moments for either beam charge,
\begin{equation}
      A_\mathrm{C}^{\cos \phi }=
          \frac { \int_0^{2\pi} d\phi \cos\phi \, \frac {d\sigma^+} {d\phi} }
                { \int_0^{2\pi} d\phi          \, \frac {d\sigma^+} {d\phi} }
        - \frac { \int_0^{2\pi} d\phi \cos\phi \, \frac {d\sigma^-} {d\phi} }
                { \int_0^{2\pi} d\phi          \, \frac {d\sigma^-} {d\phi} },
\end{equation}
is shown in figure \ref{toller} versus the missing mass $M_x$.
As expected, the asymmetry appears to be non--zero only in the exclusive region,
while in the non--exclusive region it is consistent with zero.
The asymmetry given by this formula is not due to the
real part of the DVCS amplitudes alone, but has a small
modulation from the BH amplitudes.
However, for this preliminary result this method is advantageous due to
reduced systematic uncertainties.

In summary, the beam--charge asymmetry in the hard electroproduction
of real photons has been measured for the first time. 
A sizeable asymmetry of $0.11 \pm 0.04$~(stat) $\pm 0.03$~(sys)
has been found in the exclusive region confirming the theoretical 
expectation.

\end{document}